\documentstyle[12pt]{article}

\newcommand{\beq}{\begin{equation}}
\newcommand{\eeq}{\end{equation}}

\begin{document}
\def\lag{\langle}
\def\rag{\rangle}
\clubpenalty=10000
\widowpenalty=10000

\title{Multicanonical Spin Glass
Simulations}

\author{
Bernd A. Berg$^{1,2}$  and Tarik Celik$^{1,3}$}

\vspace{.2in}
\footnotetext[1]{Supercomputer Computations Research Institute, Florida
State University, Tallahassee, FL~32306, USA.}
\footnotetext[2]{Department of Physics, The Florida State University,
Tallahassee, FL~32306, USA.}
\footnotetext[3] {On leave of absence from Department of Physics,
Hacettepe University, Ankara, Turkey.}
\maketitle

\begin{abstract}
We report a Monte Carlo simulation of the $2D$ Edwards-Anderson spin glass
model within the recently introduced multicanonical ensemble. Replica on
lattices of size $L^2$ up to $L=48$ are investigated. Once a true
groundstate is found, we are able to give a lower bound on the number of
statistically independent groundstates sampled. Temperature dependence of
the energy, entropy and other quantities of interest are easily calculable.
In particular we report the groundstate results.
Computations involving the spin glass order parameter are more tedious. Our
data indicate that the large $L$ increase of the ergodicity time is reduced
to an approximately $V^3$ power law.
Altogether the results suggest that the multicanonical
ensemble improves the situation of simulations for spin glasses and other
systems which have to cope with similar problems of conflicting
constraints.
\end{abstract}

\clearpage
\section{ Introduction}

The theoretical understanding of spin glasses (for reviews see
\cite{Bi1,book}) has remained a great challenge. In particular the low
temperature limit leaves many open questions about the effects of disorder
and frustration. For instance, it has remained controversial whether
Parisi's \cite{Paris} mean field theory provides the appropriate
description for 3D spin glasses. The attractive alternative is the droplet
model \cite{Drop}, which in turn is equivalent to a one parameter scaling
picture \cite{Bray}. The simplest spin glass system to study such questions
numerically is the Edwards-Anderson model. In its Ising version it is
described by the Hamiltonian
$$ H\ =\ - \sum_{<ij>} J_{ij} s_i s_j , \eqno(1) $$
where the sum goes over nearest neighbors and the exchange interactions
$J_{ij}=\pm 1$ between the spins $s_i=\pm 1$ are quenched random variables.
In our investigation we impose the constraint $\sum J_{ij} = 0$ for each
realization.
Despite its simplicity the model is supposed to be sufficiently realistic
to catch the physics essence correctly. Recent simulations \cite{Cara} of
the 3D model in a magnetic field support the mean field picture.
However, one may well argue that sufficiently low temperatures on
sufficiently large systems have not been reached. For previous
simulations of the Edwards-Anderson model without magnetic field
in 2D, 3D and 4D, see \cite{Ogie,Bhatt,SweWa}.

Low temperature simulations of spin glasses suffer from a slowing down
which is likely to increase with a high power law or even
exponentially fast with lattice size.
The reason is that one has to sample many independent states
separated by energy barriers which may grow with lattice size.
To illustrate the problem, let us consider
a simple ferromagnet: the 2D Ising model on a $50\times 50$ lattice.
In Figure~1 we give its magnetic probability density.
The two distinct branches below the Curie temperature
are associated with free energy valleys in configuration space, each of
which defines a (pure) thermodynamic state. The notations phases or
ergodic components are also used. At temperatures below the Curie point
the ergodicity time$^1$
\footnotetext[1]{{As will become clear in the next section, it is for the
present investigation of spin glasses more appropriate to use the term
ergodicity time $\tau^e$, instead of tunneling time $\tau^t$ which is
appropriate in the context of surface tension investigations.}}
$\tau_L^e$ increases exponentially fast with lattice size, asymptotically
like $\exp [ f^s(\beta ) L^{D-1} ]$, where $f^s$ is the surface free
energy. Therefore, on large lattices at sufficiently low temperature the
simulation of the system will, given a reasonable finite amount of
computer time, never tunnel from one phase to the other. Besides for
particular problems, like the determination of the order-order surface
free energy \cite{Bi2,our2}, this lack of tunneling does not impose a major
handicap on Ising model simulations. The reason is that the two
configuration space valleys are related by the exact symmetry
$s_i \to -s_i$ of the Hamiltonian. Exploring one valley by means of a
simulation yields also all the properties of the other one and, hence,
allows to overlook the entire system.

\begin{figure}
\vspace{11cm}
\caption[fig1]{Ising model magnetic probability density from a
$50\times 50$ lattice.}
\end{figure}

The situation is much more involved for spin glasses. For low enough
temperature the system is supposed to split off into many thermodynamic
states, separated by similar tunneling barriers as the two pure states
of the Ising model. However, unlike in the Ising model the states are not
related to each other by a symmetry of the Hamiltonian. Rather they appear
because of accidental degeneracy which in turn occurs because of randomness
and frustration of the system. For computer simulations this means that
one would like to explore many
independent configuration space valleys while keeping track of their
relative weights. The groundstate energies associated with these valleys
may or may not be degenerate, but it should be noted that even if they
are not degenerate, tunneling between the valleys would still be governed
by the energy barriers. The physics of these barriers is far less well
understood as in the ferromagnetic case. As detailed finite size scaling
(FSS) studies do not exist, it is unclear to us to what extent these
barriers depend on the system size, whereas the temperature dependence has
been investigated \cite{Bi1}.
We use the notation bifurcation temperature
(bifurcation point) for the temperature at which the spin glass
configuration space (phase transition or not) begins to split off into a
number of valleys which are well separated by energy barriers. In the
present paper we suggest that the increase of $\tau^e_L$ can be reduced
to a fairly decent power law by performing a simulation which covers in a
single ensemble a whole temperature range from well above to far below
the bifurcation point. The appropriate formulation is provided by a
generalization of the multicanonical ensemble \cite{our1}, which we
introduce in section~2. To test the approach we have performed
multicanonical simulations of the 2D Ising ferromagnet and then of
the 2D Edward-Anderson Ising spin glass model, and our numerical results
are reported in section~3. We concentrate on ground state properties
what is a kind of worst case scenario for the performance of the
multicanonical simulation.
It should be noted that Figure~1 does not exploit exact results like the
symmetry $s_i\to -s_i$, but relies on a multicanonical simulation, what
explains the slight asymmetry between the two branches.
In section~4 the multicanonical performance in comparison with other
simulation methods is evaluated in some detail. Our conclusions are
summarized in section~5. A short account of the present work has already
been reported \cite{our3}.

\section{The Multicanonical Ensemble}

Ever since the pioneering paper by Metropolis et al.\cite{Metro} most MC
simulations concentrated on importance sampling for the canonical Gibbs
ensemble. It has always been well-known, for instance \cite{Bi0},
that it is allowed to choose phase-space points according to any other
probability distribution, if it is convenient. However, a systematic
reasoning for a better than canonical choice has rarely been put forward.$^1$
\footnotetext[1]{{Notable may be microcanonical simulations, but for
the ergodicity problems on which we focus here they perform even worse
than the canonical approach does.}}
It is our suggestion that in a large class of situations,
in particular those where canonical simulations face severe ergodicity
problems, it is more efficient to reconstruct the Gibbs
ensemble from a simulation of a multicanonical ensemble \cite{our1}
than simulating it directly. In canonical simulations configurations
are weighted with the Boltzmann factor
$$
P_B (E)\ =\ \exp (-{\hat\beta} E) . \eqno(2)
$$
Here $E$ is the energy of the system under consideration, and in this paper
we use the notation $\hat\beta$ in connection with the canonical ensemble.
The resulting canonical probability density is
$$
P_c (E)\ \sim\ n(E) P_B(E), \eqno(3)
$$
where $n(E)$ is the spectral density. In order of increasing severity
problems with canonical spin glass simulations are:

\begin{itemize}

\item {\it i}) Simulations at many temperatures are needed to get an overview
         of the system.
\item
{\it ii}) The normalization in equation (3) is lost. It is
          tedious to calculate important physical quantities like the
          free energy and the entropy.
\item
{\it iii}) The low temperature  ergodicity time $\tau_L^e$ diverges
           fast with lattice size (either exponentially or with a
           high power law). The relative weights of pure states
           can only be estimated for small systems.
\end{itemize}

Let us choose an energy range $E_{\min}\le E\le E_{\max}$ and
define for a given function $\beta (E)$ the function $\alpha (E)$ by
the recursion relation (with the Hamiltonian~(1) the energy changes
in steps of 4)
$$
\alpha (E-4)\ =\ \alpha (E) +
\left[ \beta (E-4) - \beta (E) \right] E,\
\alpha(E_{\max}) = 0  . \eqno(4)
$$
The purpose of the function
$\alpha (E)$ is to give $\beta (E)^{-1}$ the interpretation of an
effective temperature.
The multicanonical ensemble \cite{our1} is then defined by weight factors
$$
P_M (E)\ =\ \exp \left[ -\beta (E) E + \alpha (E) \right] ,\eqno(5)
$$
where $\beta (E)$ is determined
such that for the chosen energy range the resulting multicanonical
probability density is approximately flat:
$$
P_{mu} (E)\ =\ c_{mu}\ n(E) P_M (E)\ \approx\ \hbox{const.} \eqno(6)
$$
In the present study we take $E_{\max} =0$ ($\beta (E) \equiv 0$ for
$E\ge E_{\max}$) and $E_{\min}=E^0$ the ground state energy of the
considered spin glass realization.

In contrast to first order phase transitions \cite{our1}, where finite
size scaling allows an accurate determination of the multicanonical
parameters from previously simulated smaller systems, the situation is
now more involved. Already changing the realization, {\it i.e.}, the quenched
random variables $J_{ij}$, for fixed lattice size leads to a situation
which requires to calculate new multicanonical parameters from scratch.
Our main  experience with such calculations is that they can be done and
that they are less problematic than one might superficially expect.
For instance, a multicanonical function $\beta (E)$ can be obtained  via
recursive MC calculations. One performs simulations with  $\beta^n (E)$,
$n=0,1,2,...$, which yield probability
densities $P^n (E)$ with medians $E^n_{\tenrm median}$. For
$E < E^n_{\min} < E^n_{\tenrm median}$ the probability density $P^n (E)$
becomes unreliable due to insufficient statistics, caused by the
exponentially fast fall-off for decreasing $E$. We start off with $n=0$
and $\beta^0 (E) \equiv 0$. The recursion from $n$ to $n+1$ reads
$$ \beta^{n+1} (E)\ =
\left\{
   \begin{array}{ll}
   	 \beta^n (E)\   \hbox{for} \ E \ge E^n_{\tenrm median};& \\
\\
   	 \beta^n (E)\ + 0.25\times\ \ln \left[ P^n(E+4)/P^n(E) \right]\\
\qquad\qquad	 \hbox{for\ \ }   E^n_{\tenrm median} > E \ge E^n_{\min}; &\\
\\
   	 \beta^{n+1} (E^n_{\min})\   \hbox{for\ \ }  E < E^n_{\min}\, .&{}
\end{array}
\right.
\eqno(7) $$
Here the $n^{\hbox{th}}$ simulation may be constrained to
$E < E^{n-1}_{\tenrm median}$ by rejecting all proposals with energy
$E > E^{n-1}_{\tenrm median}$, but one has to be careful with such bounds
in order to maintain ergodicity. The recursion is stopped for $m$ with
$E^{m-1}_{\min}=E^0$ being groundstate.

Starting with this simple approach we have explored several more
sophisticated variants. Considerable speed-ups and gains in stability could
be achieved. The CPU time spent to estimate the multicanonical parameters
was 10\% to 30\% of the CPU time spent for simulations with the final set.
Nevertheless, our determinations of the multicanonical parameters have
remained kind of unsystematic. We are not yet able to report a theoretically
sound optimized automatic procedure, although we are convinced that this
will be the final outcome.

Once the functions $\beta (E)$ and $\alpha (E)$ are fixed, the multicanonical
simulation exhibits a number of desirable features:

\begin{itemize}

\item
{\it i}) By reweighting \cite{Bau,FS} with
$\exp [-\hat\beta E + \beta (E) E - \alpha (E)]$ the canonical expectation
values
$$ {\cal O} (\hat\beta)\ =\ Z({\hat\beta} )^{-1}
\sum_E {\cal O} (E)\ n(E)\ \exp (-\hat\beta E) , \eqno(8) $$
where
$$ Z(\hat\beta )\ =\ \sum_E n(E)\ \exp (-\hat\beta E) \eqno(9) $$
is the partition function, can be reconstructed for all $\hat\beta$ in a
range $\beta_{\min} \le \hat\beta \le \beta_{\max}$, where
$\beta_{\min} = \beta (E_{\max})$ and $\beta_{\max} = \beta (E_{\min})$
follow from the requirement
$E_{\max} \ge E (\hat\beta ) \ge E_{\min}$, and $E(\hat\beta )$
follows from (8) with ${\cal O} (E) = E$. This feature
inspired the name multicanonical ensemble. With our choice $E_{\max}=0$
and $E_{\min}=E^0$ groundstate, $\beta_{\min}=0$ and
$\beta_{\max}=\infty $ follows.

\item
{\it ii}) The normalization constant $c_{mu}$ in equation (6)
follows from $Z(0)=\sum_E n(E) = 2^N$, where $N$ is the total number of
spin variables. This gives the spectral density and allows to calculate
the free energy as well as the entropy. (Remember, $\hat\beta =0$
is included in our choice of the multicanonical ensemble.)
\item
{\it iii}) We conjecture that the slowing down of canonical
low temperature spin glass simulations becomes greatly reduced.
For the multicanonical ensemble it can be argued \cite{our1}
that single spin updates cause a 1D random walk behavior of the energy $E$.
As $E_{\max} - E_{\min} \sim V$, one needs $V^2$ updating steps to cover
the entire ensemble. For first order phase transition the observed slowing
down \cite{our1,our2} was only slightly worse than this optimal behavior.
Our present MC data show more drastic modifications for spin glass
simulations.  See section~3 for further details.

\end{itemize}

To quantify our discussion of the slowing down, we have to define the
ergodicity time $\tau^e_L$. Roughly speaking it is the CPU time needed
to collect independent configurations, here for groundstates which
are our main interest. Regarding the definition of $\tau^e_L$, an
$L$-independent over-all factor is free.
For the purposes of this paper we define $\tau^e_L$ as the average number
of sweeps needed to move the energy from $E_{\max}$ to $E_{\min}$ and
back. A sweep is defined by updating each spin on the lattice once
(in the average once if random updating is implemented). We denote by
$n_{\tau}$ the number of ``tunneling'' events with respect to the
ergodicity time (suppressing now the obvious $L$ subscript).
Assuming that the correct groundstate is found, $n_{\tau}$ gives a
lower bound on the number of independent groundstates sampled.
This follows from another trivial, but remarkable property of the
multicanonical ensemble: Each time a sweep is spent at $\beta (E)\equiv 0$,
the memory of the previous Markov chain is lost entirely, and a truly
independent new series of configurations follows.
The condition $E \ge E_{\max}$ is appropriate to substitute for the
somewhat too strict constraint of an entire sweep at $\beta\equiv 0$.
As a corollary: with a disordered starting
configuration the multicanonical ensemble is immediately in equilibrium.

The energy density $e(\hat\beta )$, the specific heat $c(\hat\beta )$, the
free energy density $f(\hat\beta )$ and the entropy per spin $s(\hat\beta )$
follow in a straightforward manner by reconstructing the canonical
ensemble (8,9). In this paper we go for the extreme and concentrate on the
zero temperature $(\hat\beta\to\infty )$ limit. Let $N=L\times L$ be the
total number of spins ($N=V$ in our notation).
The groundstate energy density is $e^0 = E^0/N$,
and we obtain its entropy per spin from (8) as $s^0 = S^0/N = \ln [n(E^0)]/N$.
More complicated are calculations which aim at the spin glass order
parameter $q$. One  way to define \cite{Bi1} $q$ is as the overlap of two
statistically independent replica
$$ q(\hat\beta )\ = \ {1\over N} \sum_{i=1}^N s^1_i s^2_i \, .
   \eqno(10) $$
Here we consider equilibrium configurations with respect to the canonical
ensemble, $s^1_i$ are the spins corresponding to configurations of
replica one and $s^2_i$ are those corresponding to replica two.
Different replica have identical quenched random variables $J_{ij}$
and in practice the independence is achieved by creating uncorrelated
disordered starting configurations. The problem is now that we slow
down by at least another factor $1/V$ when we simply simulate each
replica with respect to the multicanonical ensemble. The reason is
that the configurations $s^1_i$  and $s^2_i$ will normally be at
vastly different effective temperatures $\beta (E^1)^{-1}$ and
$\beta (E^2)^{-1}$, with $q (\hat\beta )$ then being
suppressed by the reweighting procedure
at all temperatures $\hat\beta$. Here $E^1$ and $E^2$ denote the energy
replica one and two, respectively. To by-pass the problem, we decided
to constrain the multicanonical simulation of the second replica to
an appropriate energy range $E^1-\triangle E \le E^2 \le E^1+\triangle E$
and to update until $E^2$ hits the value $E^1$. In this way, we measure
a microscopic spin glass order parameter $q(E)$, which can be reweighted
without problems. In the infinite volume limit
$q(\hat\beta )\to q(E(\hat\beta ))$. The spin glass susceptibility density
$\chi_q$ and the Binder parameter are then defined as usual
$$ \chi_q = \left< q^2 \right> ~~~\hbox{and}~~~ B_q = {1\over 2}
\left[ 3 - {\left< q^4 \right> \over \left< q^2 \right>^2} \right] .
\eqno(11) $$
Again, we are mainly interested in the zero temperature results, denoted
by $\chi^0_q$ and $B^0_q$.
The constrained MC of the second replica eats the
bulk of the CPU time. In the average one needs several updating sweeps
to find $E^2=E^1$, because one has to choose $\triangle E$ generous in order
to avoid ergodicity problems. Although our procedure works, and
our results of section~3 are based on it, we do not recommend it for
future use. The better approach seems to be to use
instead of (10) an overlap function which is defined with respect to the
groundstate ensemble \cite{Bi1}. As tunneling with respect to our ergodicity
time separates truly independent groundstates, one can even avoid the second
replica entirely, while keeping the physics goals in essence unchanged.
Unfortunately, we have not yet practical experience with this approach.

To conclude this section let us comment on the connection with simulated
annealing \cite{Kirk}. A major shortcoming of simulated annealing is
that the connection with canonical equilibrium configurations gets lost. The
multicanonical ensemble overcomes this problem. The price paid is random
walk-like energy changes in contrast to more directed changes in case of
simulated annealing. Consequently, the
multicanonical ensemble is confined to more limited temperature gradients
than simulated annealing. Whether this is really a disadvantage remains
to be explored, as the smaller temperature gradients will increase
the ability of the simulation to find its way around (or in and out of)
metastable states. In simulated annealing the temperature gradients
are free parameters, whereas in our case they are fixed by the ensemble.

\section{Simulations}

All our numerical calculations were performed on the SCRI cluster of
RISC workstations. Aiming mainly at simplicity, we developed a simple
Metropolis type program. Different realizations of the spin glass
system correspond to statistically independent sets of quenched random
variables $J_{ij}$. By allowing for $J_{ij}\equiv +1$ and $J_{ij}\equiv -1$,
the program accommodates the Ising ferromagnet and anti-ferromagnet as
special cases. The multicanonical parameters are coded into an effective
action table $A(\cdot )$, new spins are proposed randomly and then
rejected with probability $(1-\min [1,\exp (A'-A)])$, where $A'$
is the effective action of the new configuration.

As an exercise and to check our code on exact results, we performed a
multicanonical simulation of the 2D Ising model with
$0\le\hat\beta < \infty$. We kept the time series of
two million sweeps and measurements on a $25\times 25$ lattice and
verified that the finite lattice
specific heat results of Ferdinand and Fisher \cite{Ferd}
are well reproduced. No difficulties are encountered with the
multicanonical ensemble when crossing the phase transition point.
To explore the possibility of zero temperature entropy
calculations, we used $Z(0)=2^{625}$ as input and obtained
$S^0 = 0.61 \pm 0.09$ for the total groundstate entropy. This
corresponds to an estimate of $1.84 \pm 0.17$ groundstates, {\it i.e.},
within statistical errors we are in agreement with two. Using
$Z(0)=2^{2500}$ and a time series of four million sweeps on a $50\times 50$
lattice we obtained $2.07 \pm 0.22$ for the number of groundstates.
Figure 1 is obtained from the simulation of this lattice.

After this test we turned immediately to the 2D Edwards-Anderson spin
glass. On lattices of size $L=4$, 12, 24 and 48 we performed multicanonical
simulations. Up to $L=24$ we investigated ten different realizations per
lattice and, due to CPU time constraints, we considered only five
realizations for the $L=48$ lattice. To study the simulation method,
a rather small numbers of realizations is sufficient and to some extent
desirable, as at this stage each single realization still deserves some
individual attention. The
multicanonical energy distribution for the second of our $L=48$ realizations
is depicted in Figure~2. The fall-off for $-e < 0$ is like that of the
canonical distribution at $\hat\beta =0$. For $0\le -e < -e^0$ an
impressive flatness (about 800 energy entries on the lattice under
consideration) is quickly achieved by the recursion (7). Close to the
groundstate some difficulties are encountered on which we comment later.
It should be understood that deviations from the desired constant
behavior (6) do only influence the statistical error bars, but not the
estimates themselves. Therefore, such deviations do not pose problems as
long as they can be kept within reasonable limits of approximately one order
of magnitude. In Figure~3 the function $\beta (E)$ is given versus
$E$ for the same realization as used for Figure~2.

\begin{figure}
\vspace{9cm}
\caption[fig2]{Multicanonical energy density distribution ($L=48$ lattice,
realization 2).}
\end{figure}

\begin{figure}
\vspace{9cm}
\caption[fig3]{Multicanonical $\beta (E)$ function ($L=48$ lattice, realization
2).}
\end{figure}
Tables 1--4 give an overview of our numerical results. All entries are
as introduced in the previous sections. For $\beta_{\max}$ we take
$\beta (E^0)$, where it should be noted that due to
our computational procedure $\beta (E)$ is a noisy function. The final
mean values and their error bars are obtained by combining the
results from the different realizations. Different realizations are
statistically independent and enter with equal weights. The final
error bar is enlarged by a Student multiplicative factor, such
that the probability content of two standard deviations is
Gaussian (95.5\% ). The dependence of the
estimates on the realizations is fairly strong. Optimal use of CPU
time is made by calculating estimates on the realizations
with identical variance than the one obtained by the
dependence on the realizations. Therefore, for the smaller lattices it
would have been more efficient to simulate a larger number of realizations
with less CPU time spent on each. Towards larger systems one expects
self-averaging of energy and entropy with respect to the quenched random
variables $J_{ij}$, and a decrease of the variance between
different realizations like $1/N$. Our data are consistent with such a
behavior. There is no self-averaging expected and found for quantities
which relate to the spin glass order parameter.

\begin{table}
\begin{small}
\begin{center}
\begin{tabular}{|c|c|c|c|c|c|c|c|}                    \hline
\# & $\beta_{\max}$ & $\tau^e_{4}$ & $n_{\tau}$ & $e^0$ & $s^0$
                             & $\chi_q^0$ & $B^0_q$ \\  \hline
 1  & 0.80 & 34 & 46786 &$-$1.2500 & 0.1871 (3) & 0.4747 (11) & 0.8060 (12)
 \\ \hline
 2  & 0.78 & 29 & 55177 &$-$1.2500 & 0.1989 (4) & 0.3762 (17) & 0.5847 (38)
 \\ \hline
 3  & 0.84 & 40 & 40261 &$-$1.2500 & 0.1732 (4) & 0.6155 (06) & 0.9471 (02)
 \\ \hline
 4  & 0.94 & 41 & 39536 &$-$1.2500 & 0.1556 (4) & 0.4870 (23) & 0.7210 (27)
 \\ \hline
 5  & 0.82 & 43 & 37066 &$-$1.2500 & 0.1555 (3) & 0.4672 (15) & 0.7117 (20)
 \\ \hline
 6  & 0.81 & 25 & 63873 &$-$1.2500 & 0.2161 (3) & 0.3434 (10) & 0.6035 (18)
 \\ \hline
 7  & 0.41 & 22 & 71464 &$-$1.2500 & 0.2472 (3) & 0.3507 (07) & 0.7270 (12)
 \\ \hline
 8  & 0.73 & 41 & 39444 &$-$1.2500 & 0.1560 (4) & 0.4447 (23) & 0.6370 (36)
 \\ \hline
 9  & 0.52 & 35 & 46074 &$-$1.5000 & 0.1117 (4) & 0.8020 (04) & 0.9885 (01)
 \\ \hline
10  & 0.82 & 43 & 36875 &$-$1.2500 & 0.1544 (3) & 0.4723 (18) & 0.7178 (19)
 \\ \hline
Mean &    0.75 &   35.3 & &$-$  1.275 &     0.178 &     0.483 &     0.742
 \\ \hline
Error&$\pm$ .06~~~&$\pm$ 2.8~~~& &$\pm$ 0.029 &$\pm$ .014~~~&$\pm$ .051~~~&
      $\pm$ .049~~ \\ \hline
\end{tabular}
\caption[tab1]{{\em $L=4$ results. Each realization relies on 1,600,000
sweeps.}}
\end{center}
\end{small}
\end{table}

\begin{table}
\begin{small}
\begin{center}
\begin{tabular}{|c|c|c|c|c|c|c|c|}                    \hline
\# & $\beta_{\max}$ & $\tau^e_{12}$ & $n_{\tau}$ & $e^0$ & $s^0$
                             & $\chi_q^0$ & $B^0_q$ \\  \hline
 1  & 1.53 & 2647 & 604* &$-$1.4167 & 0.0810 (06) & 0.465 (06) & 0.918 (04)
 \\ \hline
 2  & 1.71 & 2871 & 138~~~&$-$1.3889 & 0.0943 (08) & 0.379 (17) & 0.837 (27)
 \\ \hline
 3  & 1.52 & 3603 & 111~~~&$-$1.3333 & 0.1005 (09) & 0.434 (15) & 0.859 (18)
 \\ \hline
 4  & 1.39 & 3350 & 119~~~&$-$1.4167 & 0.0754 (11) & 0.369 (25) & 0.742 (30)
 \\ \hline
 5  & 1.43 & 1406 & 283~~~&$-$1.3611 & 0.1315 (05) & 0.374 (08) & 0.909 (06)
 \\ \hline
 6  & 1.44 & 1453 & 274~~~&$-$1.3056 & 0.1328 (06) & 0.188 (10) & 0.503 (38)
 \\ \hline
 7  & 1.35 & 1959 & 204~~~&$-$1.3889 & 0.1001 (09) & 0.447 (11) & 0.887 (09)
 \\ \hline
 8  & 1.29 & 1659 & 241~~~&$-$1.4167 & 0.0873 (06) & 0.406 (15) & 0.836 (15)
 \\ \hline
 9  & 1.48 & 5296 & 301* &$-$1.3889 & 0.0696 (06) & 0.631 (05) & 0.968 (03)
 \\ \hline
10  & 1.53 & 1833 & 873* &$-$1.3333 & 0.1176 (03) & 0.325 (09) & 0.727 (14)
 \\ \hline
Mean &    1.47&      2607 & &$-$  1.375 &     0.099 &     0.405 &     0.819
 \\ \hline
Error&$\pm$ .04~~~&$\pm$ 450~~~& &$\pm$ 0.015 &$\pm$ .008~~~&$\pm$ .041~~~&
      $\pm$ .049~~ \\ \hline
\end{tabular}
\caption[tab2]{{\em $L=12$ results. Each realization relies on 400,000 sweeps,
the data points marked by * have four times this statistics.}}
\end{center}
\end{small}
\end{table}

\begin{table}
\begin{small}
\begin{center}
\begin{tabular}{|c|c|c|c|c|c|c|c|}                    \hline
\# & $\beta_{\max}$ & $\tau^e_{24}$ & $n_{\tau}$ & $e^0$ & $s^0$
                             & $\chi_q^0$ & $B^0_q$ \\  \hline
 1  & 2.00 &  40696 & 39 &$-$1.3403 & 0.1135 (05) & 0.18 (04) & 0.63 (11)
 \\ \hline
 2  & 2.45 & 391626 &  3 &$-$1.3958 & 0.0684 (20) & 0.70 (02) & 1.00 (04)
 \\ \hline
 3  & 1.87 & 145092 & 11 &$-$1.4167 & 0.0789 (07) & 0.27 (06) & 0.61 (15)
 \\ \hline
 4  & 2.31 & 338000 &  2 &$-$1.4028 & 0.0645 (18) & 0.67 (09) & 0.99 (01)
 \\ \hline
 5  & 1.94 &  97587 & 16 &$-$1.3681 & 0.0851 (06) & 0.39 (04) & 0.87 (04)
 \\ \hline
 6  & 1.97 & 159069 & 10 &$-$1.3750 & 0.0921 (10) & 0.23 (14) & 0.62 (25)
 \\ \hline
 7  & 1.97 &  67809 & 23 &$-$1.3819 & 0.0841 (04) & 0.30 (06) & 0.70 (11)
 \\ \hline
 8  & 1.78 & 204749 &  5 &$-$1.4028 & 0.0651 (19) & 0.66 (03) & 0.99 (01)
 \\ \hline
 9  & 2.20 & 173304 &  9 &$-$1.3958 & 0.0868 (16) & 0.46 (04) & 0.93 (04)
 \\ \hline
10  & 2.73 & 319574 &  4 &$-$1.4028 & 0.0745 (34) & 0.20 (20) & 0.43 (58)
 \\ \hline
Mean &    2.12& 193750 & &$-$1.388 &  0.081 &     0.406 &     0.777
 \\ \hline
Error&$\pm$ .11~~~&$\pm$ 43820~~~& &$\pm$ 0.008 &$\pm$ .006~~~&$\pm$ .077~~~&
      $\pm$ .076~~ \\ \hline
\end{tabular}
\caption[tab3]{{\em $L=24$ results. Each realization relies on 1,600,000
sweeps.}}
\end{center}
\end{small}
\end{table}

\begin{table}
\begin{small}
\begin{center}
\begin{tabular}{|c|c|c|c|c|c|c|c|}                    \hline
\# & $\beta_{\max}$ & $\tau^e_{24}$ & $n_{\tau}$ & $e^0$ & $s^0$
                             & $\chi_q^0$ & $B^0_q$ \\  \hline
 1  & 2.13 &  425705 & 2 &$-$1.4132 & 0.0781 (5) & 0.050 (57) & 0.70 (33)
 \\ \hline
 2  & 2.16 &  992676 & 4 &$-$1.4063 & 0.0786 (3) & 0.043 (21) & 0.57 (38)
 \\ \hline
 3  & 2.34 & 2464240 & 2 &$-$1.4080 & 0.0801 (3) & 0.141 (08) & 0.94 (04)
 \\ \hline
 4  & 2.33 & 2004270 & 2 &$-$1.3924 & 0.0812 (8) & 0.079 (79) & 0.74 (36)
 \\ \hline
 5  & 2.12 & 1399679 & 2 &$-$1.3767 & 0.0914 (9) & 0.098 (06) & 0.86 (14)
 \\ \hline
Mean& 2.22 & 1457315 &   &$-$1.399  & 0.082      & 0.084 &     0.76
 \\ \hline
Error&$\pm$ .07~~~&$\pm$ 516925~~~& &$\pm$ 0.010 &$\pm$ .004~~~&$\pm$ .025~~~&
      $\pm$ .10~~ \\ \hline
\end{tabular}
\caption[tab4]{{\em $L=48$ results. The realizations rely on 4,000,000 $-$
              6,400,000 sweeps.}}
\end{center}
\end{small}
\end{table}

For the $L=4$ lattice we also calculated the exact results by enumeration
of all $2^{16}$ configurations per realization. As expected, the
groundstate energies reported in Table~1 are identical with the exact
values. For all other quantities of Table~1 we find reasonable agreement
within the statistical errors. The about $2\sigma$ discrepancy of our
$B^0_q$ ($L=4$) with \cite{Bhatt} should therefore be interpreted as a
statistical fluctuation.

\begin{figure}
\vspace{8.5cm}
\caption[fig4]{Ergodicity times versus lattice size on a double log scale.}
\end{figure}

In Figure 4 we plot the ergodicity time versus lattice size $L$ on
a log--log scale. The data are consistent with a straight line fit
($Q$ denotes the goodness of fit \cite{Recipes}), which gives the
finite size behavior
$$ \tau^e_L\ \sim\ L^{4.4 (3)} ~~~\hbox{sweeps}. \eqno(12)$$
In CPU time this corresponds to a slowing down $\sim V^{3.2 (2)}$. It should
be remarked that a fit of form $\tau^e_L \sim \exp (c L)$ results
in a completely unacceptable goodness of fit $Q < 10^{-6}$. Still,
the behavior (12) is by an extra volume factor worse than the close to
optimal performance we hoped for. To understand the reasons for the
rather high power, we first focused on the acceptance rate of the
Metropolis updating, which is depicted in Figure~5 for our various
lattice sizes. In all cases we find a drop towards about 10\% in the
ground state vicinity. This lattice size dependence is too weak to provide
on its own a sufficient explanation for the extra volume factor encountered
in (12). Therefore, we investigated next which new states are really accepted.
We find that in the vicinity of the groundstate most accepted proposals do
not change the energy, whereas in the disordered region the energy is
changed in a majority of the accepted cases. Let us denote by
$P_{\tenrm move}$ the probability that an accepted new state has a
different energy then the previous state. Table~5 summarizes the groundstate
$P^0_{\tenrm move}$ probabilities versus lattice size. The observed decrease
is conjectured to explain the encountered extra volume factor. The energy
random walk picture is still correct, but the Metropolis algorithm moves
more slowly with increasing lattice size. It may be that rather
straightforward heat bath consideration with respect to the lattice
configuration as a whole could fix the problem, and we intend to approach
this problem in future work. As small $P_{\tenrm move}$ probabilities are
limited to the groundstate neighborhood, they provide also a natural
explanation for the difficulties which we encountered there when
determining the multicanonical parameters.

\begin{figure}
\vspace{8.5cm}
\caption[fig5]{Multicanonical Metropolis acceptance rates.}
\end{figure}

\begin{table}
\begin{center}
\begin{tabular}{|c|c|}                    \hline
   $L$  &  $P^0_{\tenrm move}$   \\  \hline
    4   &  $0.392 \pm 0.026$  \\ \hline
   12   &  $0.196 \pm 0.018$  \\ \hline
   24   &  $0.096 \pm 0.021$  \\ \hline
   48   &  $0.196 \pm 0.143$  \\ \hline
\end{tabular}
\caption[tab5]{{\em Metropolis ``move'' probability versus lattice size.}}
\end{center}
\end{table}

\begin{figure}
\vspace{16cm}
\caption[fig6]{Energy density and entropy per spin versus $\beta$ (from the
$L=48$ lattices).}
\end{figure}

Figure 6 illustrates the possibility to calculate canonical expectation
values for all $\hat\beta \ge 0$. Energy density $e = e(\hat\beta )$ and
entropy per spin $s = s(\hat\beta )$ for $L=48$ are depicted
$(0\le \hat\beta \le 3)$. The indicated error bars are with respect
to the five different realizations. It should be noted that error
bars at subsequent $\hat\beta$ values are highly correlated, because
the underlying configurations are identical, the only difference being
in the re-weighting. For small $\hat\beta$ the error bars are not visible
on the scale of Figure~6 (the horizontal bar is not an error).

\begin{figure}
\vspace{16cm}
\caption[fig7]{FSS estimate of the infinite volume groundstate energy density
and entropy per spin.}
\end{figure}

We estimate the infinite volume groundstate energy and entropy from
FSS fits of the form $f^0_L = f^0_{\infty} + c/V$.
These fits are depicted in Figures 7. Our groundstate energy
estimate $e^0 = -1.394 \pm 0.007$ is consistent with the previous
MC estimate \cite{SweWa} $e^0 = -1.407 \pm 0.008$ as well as with the
transfer matrix result \cite{CMc} $e^0 = -1.4024 \pm 0.0012$. Our
groundstate entropy estimate $s^0 = 0.081 \pm 0.004$ is also consistent
with the MC estimate \cite{SweWa} $s^0 = 0.071 \pm 0.007$, but barely
consistent with the more accurate transfer matrix result
\cite{CMc} $s^0 = 0.0701 \pm 0.005$.
Still, a larger statistical sample and a careful study of
systematic error sources would be needed to claim that there is a
significant discrepancy. Our results could be improved by exploiting
the high temperature expansion of the entropy, as it was done in
\cite{SweWa}. However, this would be against the spirit of this paper
which is to explore the possibilities and limits of multicanonical spin
glass simulations. It should be noted that the reported groundstate
entropy value translates, even for moderately sized systems, into
large numbers of distinct groundstates. For instance $s^0=0.075$
implies the following approximate groundstate numbers:
$4.9\times 10^4\ (L=12)$, $5.8\times 10^{18}\ (L=24)$, and
$1.1\times 10^{75}\ (L=48)$.

Our results relying on groundstate data for the spin glass order
parameter are of less satisfactory quality. Due to lack of self-averaging
one has to sample many groundstates.
For $L\ge 12$ we scaled our CPU time approximately with $\sim V^2$,
and not $\sim V^3$ as would (presently) be required to keep the number of
independent groundstates visited approximately constant. The number
of tunneling events $n_{\tau}$ decreases with lattice size, and for $L=48$
our largest number is only $n_{\tau}=4$ (achieved by our favorite
realization). Although $n_{\tau}=4$ is a too small number for serious
statistical conclusions, this case serves well to illuminate the
problems as well as the achievement of the multicanonical simulation.
In analogy to Figure~1, we give in Figure~8 the spin glass order parameter
distribution of this realization. Instead of two branches we find now four.
However, there is an important difference: The number of tunneling events
may now be small in comparison to the number of expected branches. In the
Ising model we know that there are only two branches of the magnetization,
and for the $50\times 50$ lattice on which Figure~1 relies the number of
tunneling events was
$n_{\tau}=118$. Therefore, in the Ising model we could assemble reasonable
statistics for each branch. Now, we may have to cope with
collecting a representative sample from a huge number of all branches.
Depending on the number of tunneling events Figure~8 may look different,
the number of groundstate ``fingers'' roughly proportional
to the number of tunneling events. Figure~9 shows the groundstate state
time series for the spin glass order parameter. The time $t_0$ counts
only those sweeps which emerge in a groundstate and $q_0$ is the
corresponding overlap of our independent replica. Of our
four million data little more than 1\% (precisely 4,419) belong to the
groundstate energy. These 1\% decompose into four segments. Within each
segment the data are highly correlated, but the segments themselves are
statistically independent, because in-between the system tunneled each time
all the way back into the completely disordered $\beta\equiv 0$ region.
Additionally, the time series moved within each section occasionally out
of the groundstate. This gives some amount of independence to the data
within each segment. The numbers over each segment give the corresponding
ranges in the total time series of our four million sweeps, and the sum of
these ranges exceeds of course 4,419. For $L=48$ the $\chi_q^0$ and $B_q^0$
estimates suffer from insufficient statistics. Entire jackknife bins
with $q_0 \approx 0$ exist, and error bars which may exceed the signal
are implied. Having these limitations in mind, the smallness of our
$\chi_q^0$ data seems still to contradict the scaling
$\chi^0_q \sim L^{-0.2}$ \cite{Bhatt}  at the phase transition point
$T=0$. After improving the simulation method (getting
rid of the constrained MC), it will be desirable to investigate a large
number of $L=48$ realizations with high statistics.

\begin{figure}
\vspace{9cm}
\caption[fig8]{Spin glass order parameter distribution ($L=48$ lattice,
realization 2).}
\end{figure}

\begin{figure}
\vspace{9cm}
\caption[fig9]{Multicanonical simulation, time series for spin glass order
parameter
ground state measurements ($L=48$ lattice, realization 2).
}
\end{figure}

\section{Evaluation of the Performance}

It is not entirely straightforward to compare multicanonical and
standard simulations. For instance autocorrelation times of
multicanonical simulations come out short due to the triviality that the
simulation spends most of its time at rather small effective $\beta $
values. Therefore, autocorrelations are not well suited. Our ergodicity
time, the average number of sweeps to find truly independent groundstates,
is a more useful quantity. When relying on it, {\it i.e.}, concentrating on
groundstate properties alone, we should remember that this pushes the
multicanonical simulation to its extreme. This way of calculating
groundstate properties gives for free all properties in-between, from
$\hat\beta =0$ on. If a phase transition occurs, its study is also included
(we noticed no particular difficulties in the 2D Ising model). For
instance in the 3D Ising spin glass the canonical slowing down at
$\hat\beta_c$ \cite{Bi1} is already worse than the one given by our equation
(12).

Although the slowing down (12) is severe, it seems to provide an important
improvement when compared with the slowing down which canonical
simulations encounter for temperatures below the bifurcation temperature.
For $L\ge24$ canonical simulations \cite{Bi1,Bhatt} are unable to
equilibrate the systems at the $\beta_{\max}$ values reported in our
tables since the relaxation time is by far too long. Presumably due to
this fact the literature focused on other questions and does not provide
detailed FSS investigation of relaxation times. A rough
estimate of the canonical ergodicity time may be \cite{Bi1}
$$ \tau^e_{\hbox{canonical}}\ \sim\ \exp \left( C \hat\beta - C' \right)
\, .\eqno(13) $$
In this equation the scaling with $L$ may be hidden in the $L$
dependence of our $\beta_{\max}$ values, which is argued to be divergent
like $\beta_{\max} \sim \ln (V)$, and this line of reasoning gives a
slowing down of the canonical algorithm like $V^{15}$. With
$\beta_{\max}=2.12$ (our $L=24$ case) equation~(13) leads to a
canonical ergodicity time of order $10^8-10^9$ when the missing constant
is assumed to be of order one, whereas our $\tau^e_{24}$ is about
$2\times 10^5$.

To get a more direct handle on these problems, we performed canonical
simulations at $\hat\beta =1.4$ for our $L=12$ realizations and at
$\hat\beta =1.9$ for our $L=24$ realizations ($\hat\beta$ values
somewhat lower than the corresponding $\beta_{\max}$ averages of
Tables~$1-4$ seem to be kind of optimal for canonical groundstate
investigations). In each case  two independent replicas were simulated
and the spin glass order parameter was calculated ala (10). As our
ergodicity time $\tau_e$ and the corresponding number of
tunneling events $n_{\tau}$ are nonsense for canonical
simulations, we replace $n_{\tau}$ by a definition which allows
to compare canonical and multicanonical simulations. Monitoring
the groundstate time series $q^0(t)$ suggests that its number of
sign flips $n_f$ is a suitable replacement for $n_{\tau}$. To
disregard small fluctuations around zero, we introduce a cut-off
range: A sign flip is counted when $q^0(t)$ tunnels form one side
of the cut-off range to the other. We found that $\pm 0.3472$ gives
an appropriate range, with $L=12$ this is $\pm 50$ for $q^0 V$.
According to \cite{Bhatt} one may want to scale with $L^{1.8}$ instead
of $V=L^2$, but the difference would not matter for our present purposes.

Let us first report the canonical $L=12$ results. We perform an identical
number of sweeps as for the multicanonical simulations of Table~2. For
each realization the canonical simulation finds precisely the groundstate
energy reported in Table~2. In Table~6 we compare multicanonical and
canonical flip rates. For the multicanonical case we notice that
$n_f$ is smaller than $n_{\tau}$ reported in Table~2. This is due to
the fact that subsequent, independent groundstate configurations
agree with probability 1/2 on the sign of $q^0 (t)$. For all realizations
the canonical flip rate is smaller than the multicanonical. The actual
ratio depends strongly on the realization. This effect can be fairly
dramatic. In Figure~10 we consider realization \#~5 and compare
its multicanonical with its canonical $q^0(t)$ time series. For the
canonical simulation the time series never flips!
Depicted are those values of $q$
for which both replica are in a groundstate, and this is the case
for more than a quarter of the entire time series (125,000 of
400,000 sweeps). In contrast to this the multicanonical simulations
spent only 8,000 sweeps, {\it i.e.}, 2\%  of its statistics, in the
groundstate, but it tunnels often (179 flips). Going now to replica
\#~6,depicted in Figure~11, the multicanonical performance is
very similar, but the canonical simulation has now also little
difficulties and flips 81 times. To estimate the
over-all improvement factor of the multicanonical simulations is
not possible because of the two cases with $n_f =0$. If we replace
these zeros by one, we obtain $46\pm 22$, what should be considered
as some kind of lower bound.

\begin{table}
\begin{small}
\begin{center}
\begin{tabular}{|c|c|c|c|c|}                                          \hline
\#&Muca $(L=12)$&Cano $(L=12)$&Muca $(L=24)$&Cano $(L=24)$ \\ \hline
 1  & 358*   & ~11*        & 76      &  7     \\ \hline
 2  &  80~   & ~18~~       & ~4      &  0     \\ \hline
 3  &  97~   & ~29~~       & 20      &  1     \\ \hline
 4  &  61~   &  ~2~        & ~4      &  0     \\ \hline
 5  & 179~~  &  ~0~        & 44      &  0     \\ \hline
 6  & 151~~  & ~81~~       & ~4      &  1     \\ \hline
 7  & 119~~  &  ~0~        & 52      &  0     \\ \hline
 8  & 147~~  &  ~2~        &  4      &  0     \\ \hline
 9  & 217*   & ~17*        & 16      &  0     \\ \hline
10  & 483*   & 112*        &  8      &  0     \\ \hline
\end{tabular}
\caption[ab6]{{\em Number of flips $n_f$ (defined in the text) for
multicanonical (Muca) and canonical (Cano) simulations on $L=12$ and $L=24$
lattices. The $L=12$ data points marked with * have four times more statistics
than the other $L=12$ data points. }}
\end{center}
\end{small}
\end{table}

\begin{figure}
\vspace{16cm}
\caption[fig10]{(a)Multicanonical versus (b)canonical simulation: Time series
for spin glass order parameter ground state measurements ($L=12$ lattice,
realization 5).}
\end{figure}

\begin{figure}
\vspace{16cm}
\caption[fig11]{(a)Multicanonical versus (b)canonical simulation: Time series
for spin
glass order parameter ground state measurements ($L=12$ lattice,
realization 6).}
\end{figure}

For $L=24$ we performed canonical simulations with four times
the multicanonical statistics of Table~3 ,
{\it i.e.}, 6,400,000 sweeps per replica. Nevertheless we
obtain zero tunneling events for most of the realizations. In view of
the ergodicity time estimated by our previous consideration, this is
no surprise. By these zeros any attempt to estimate the $L=24$
multicanonical improvement directly is rendered hopeless: One would have
to perform canonical simulations until at least one flip is obtained
for each replica. To do so would be an obvious waste of computer
resources. The superiority of the multicanonical approach is already
clearly established by the $L=12$ results.

When one is only interested in groundstate properties, minimization
algorithms have to be compared. As a method simulated annealing
\cite{Kirk} stands out because of its generality, although there are
more efficient algorithms for special cases, which should be used
when appropriate. One problem of simulated annealing is the dependence
of the results on the cooling rate $r=-\triangle T / \hbox{sweeps}$.
True groundstates are only encountered for $r$ sufficiently small.
The $r$ dependence of groundstate energies was investigated in \cite{Grest},
and for our model the behavior
$$ e (r)\ =\ e^0 + c r^{1\over 4} ~~\hbox{with}~~ c \approx 0.5 \eqno(14) $$
is indicated. To find a true groundstate, one has to reduce
$[e(r) - e^0]$ to the order $1/V$. Assuming that the constant in
equation (14) is volume independent (only the lattice size $100\times 100$
was considered in \cite{Grest}), this translates to
$$ \hbox{sweeps}\ \sim\ V^4 = L^8 , \eqno(15) $$
far worse than our equation (12). This result is kind of amazing as
the multicanonical ensemble has eliminated directed cooling and is
nevertheless more efficient. If one does not insist on true
groundstates one can relax the condition $[e(r) - e^0] \sim 1/V$.
For instance, any behavior $[e(r) - e^0] \to 0$ with $L\to\infty$
would still give the correct density, and simulated annealing would
slow down far less dramatic then according to (15). On the other hand,
this would also imply a less stringent multicanonical simulation, and
the final outcome of such a comparison is unclear. To compare the
performance on lattices of fixed sizes is more subtle than just fixing
the constant in equations (12) and (15). The reason is that the
multicanonical simulations normally finds more than one groundstate within
one tunneling period. This will amount to an extra factor in favor of the
multicanonical simulation, which can be determined by calculating the
integrated autocorrelation time for the groundstate series. Presently
we did not attempt this, because at least for our $L=48$ realizations the
statistics are insufficient for such an enterprise.

A comparison with the replica MC algorithm \cite{SweWa} is even less
clear cut. The obtained estimates of the groundstate energy and entropy are
in accuracy similar to ours. As one has to simulate many replica at
many $\hat\beta$-values a direct comparison is impossible. Clearly, the
results reported on slowing down are much more promising than ours
for the ergodicity time.
In particular, with our present implementation we would be unable to
equilibrate a large $128 \times 128$ system. On the other hand, to our
knowledge the replica MC approach has never been applied to the 3D
Edwards-Anderson model, and one may well encounter difficulties.
In contrast, 3D multicanonical simulations are straightforward. In fact
the dimension is just a parameter in all our computer programs and we
have already carried out various 3D test runs.

Let us address the question of algorithms from a more general perspective.
With a Metropolis type implementation our optimum performance will be
bounded from below by a slowing down $\sim V^2$ in CPU time,
and we are outperformed
by any algorithm which can do better than this. For a number of important,
but often highly specialized, applications such better algorithms exist
and should be used. Cluster algorithms are presumably the most noticeable
case of specialized high performance algorithms. The main advantage of
the multicanonical ensemble is its generality. With this respect our
method resembles simulated annealing \cite{Kirk},
while clearly avoiding some of its
disadvantages. Finally, it should be stressed that the multicanonical
ensemble is an ensemble and not an algorithm. One may find better algorithms
than conventional Metropolis updating to simulate this new ensemble. To try
a combination with cluster algorithms is certainly an attractive idea.

\section{Conclusions}

Simulations of the multicanonical ensemble open new perspectives for
a wide range of applications. The present paper shows that multicanonical
spin glass simulations are feasible and some results are very encouraging.
Still, a number of technical details have remained tedious. Our believe is
that by gaining more experience with the multicanonical ensemble more
efficient implementations of many details (like determining the parameters)
will emerge. We hope that the present paper will stimulate further
investigations. By no means we do claim that simulations of the
multicanonical ensemble are already well understood.

In our present implementation we find a power law slowing down $\sim V^3$
in CPU time. Quantitative comparisons with standard simulations are kind of
difficult, as in awareness of the ergodicity problems there aims have
been different from ours. In addition, a fair comparison has to
take into account that besides groundstates our multicanonical
simulation covers the entire range from $\hat\beta =0$ on. This
amounts to an additional factor of at least $\sqrt{V}$ in favor of the
multicanonical ensemble. Even without doing so, the improvement of the
slowing down of canonical simulations is remarkable. We were able
to equilibrate the system at $\beta$ values which are simply
inaccessible to canonical simulations because of relaxation problems.
When we constrain our attention just to the problem of identifying true
groundstate configurations in the limit of large systems, our performance
seems to be superior to that of simulated annealing \cite{Kirk,Grest}.
Both methods share that they can address a very general range of
applications, but a major advantage of the multicanonical ensemble is
that the relationship to the equilibrium canonical ensemble remains
exactly controlled.
As we talk about a new ensemble (in contrast to just an algorithm), further
improvements seems to be possible through combination with algorithms
which are more efficient than simple Metropolis \cite{Metro} updating.
In particular the possibility of exploiting cluster ideas \cite{SweWa}
could be explored.

Qualitatively our results make clear that the
multicanonical approach is certainly a relevant enrichment of
the options one has with respect to spin glass simulations.
The similarities of spin glasses to other problems with
conflicting constraints \cite{Kirk} suggest that multicanonical
simulations may be of value for a wide range of investigations:
optimization problems like the travelling salesman, neural networks,
protein folding, and others.
Multicanonical simulations of the 3D Edwards-Anderson model may eventually
shed new light on the questions whether the model exhibit mean field-like
behavior or some kind of droplet picture applies \cite{Paris,Drop,Bray,Cara}.
In the later scenario one would expect that a single valley dominates the
others in the thermodynamic limit, what simply means it leads down to lower
energies. On the other hand in the mean field scenario one would expect
groundstate degeneracy over many valleys. It seems that previous numerical
work on this question is somewhat inconclusive as sufficiently low
temperatures could not be reached without destroying thermodynamic
equilibrium.

\section*{Acknowledgements}
We would like to thank Ulrich Hansmann and Thomas Neuhaus for numerous
useful discussions. Further, we are indebted to Philippe Backouche,
Norbert Schultka, and Unix\_Guys for valuable help.
T.C. was supported by TUBITAK of Turkey and likes to thank Joe Lannutti
and SCRI for the warm hospitality extended to him.
Our simulations were performed on the SCRI cluster of fast RISC
workstations.

This research project was partially funded by the the
Department of Energy under contracts DE-FG05-87ER40319,
DE-FC05-85ER2500, and by the NATO Science Program.
\vspace{1cm}

After submitting this paper we became aware of a preprint by
Marinari and Parisi [23] which explores similar ideas for the
3D random Ising model.

[23] E.Marinari and G.Parisi, preprint, ROM2F-92-06.


\begin{thebibliography}{99}

\bibitem{Bi1} K. Binder and A.P. Young, Rev. Mod. Phys. 58 (1986) 801.

\bibitem{book} K.H. Fisher and J.A. Hertz, {\it Spin Glasses}, Cambridge
                  University Press 1991.

\bibitem{Paris} G. Parisi, Phys. Rev. Lett. 43 (1979) 1754;
                J. Phys. A13 (1980) 1101.

\bibitem{Drop} D.A. Fisher and D.A. Huse, Phys. Rev. B38 (1988) 386.

\bibitem{Bray} A.J. Bray and M.A. Moore, in {\it Heidelberg Colloquium on
               Glassy Dynamics}, edited by J.L. van Hemmen and I. Morgenstern,
               Lecture Notes in Physics, Vol. 275, Springer 1987.

\bibitem{Cara} S. Caracciolo, G. Parisi, S. Patarnello and N. Sourlas,
               Europhysics Letters 11 (1990) 783.

\bibitem{Ogie} A.T. Ogielski, Phys. Rev. B32 (1985) 7384.

\bibitem{Bhatt} R.N. Bhatt and A.P. Young, Phys. Rev. B37 (1988) 5606.

\bibitem{SweWa} R.H. Swendsen and J.-S. Wang, Phys. Rev. B38 (1988) 4840.

\bibitem{Bi2} K. Binder, Phys. Rev. A25 (1982) 1699.

\bibitem{our2} B. Berg, U. Hansmann and T. Neuhaus, preprint,
               FSU-SCRI-91-125.

\bibitem{our1} B. Berg and T. Neuhaus, Phys. Lett. B267 (1991) 249;
               Phys. Rev. Lett. 68 (1992) 9.

\bibitem{our3} B. Berg and T. Celik, preprint, SCRI-92-58.

\bibitem{Metro} N. Metropolis, A.W. Rosenbluth, M.N. Rosenbluth,
                A.H. Teller and E. Teller, J.~Chem. Phys. 21 (1953) 1087.

\bibitem{Bi0} K. Binder, in {\it Phase Transitions and Critical Phenomena},
              edited by C.~Domb and M.S. Green, Vol.5B, Academic Press,
              NY 1976.

\bibitem{Bau} B. Baumann and B. Berg, Phys. Lett. 164B (1985) 131;
              B.~Baumann, Nucl. Phys. B285 (1987) 391.

\bibitem{FS} A.M. Ferrenberg and R.H. Swendsen, Phys. Rev. Lett. 61 (1988)
             2635; erratum 63 (1989) 1658; and references given in the
             erratum.

\bibitem{Kirk} S. Kirkpatrick, C.D. Gelatt and M.P. Vecchi, Science 220
               (1983) 671.

\bibitem{Ferd} A.E. Ferdinand and M.E. Fisher, Phys. Rev. 185 (1969) 832.

\bibitem{Recipes} W.H. Press, B.P. Flannery, S.A. Teukolsky and
                  W.T. Vetterling, {\it Numerical Recipes}, Cambridge
                  University Press 1988.

\bibitem{CMc} H.-F. Cheng and W.L. McMillan, J. Phys. C16 (1983) 7027.

\bibitem{Grest} G.S. Grest, C.M. Soukoulis and K. Levin, Phys. Rev. Lett.
                56 (1986) 1148.

\end{thebibliography}
\end{document}